\documentclass[letter, scriptaddress, twocolumn,  prd, showpacs,showkeys]{revtex4-1}

	\usepackage{amsmath}
        \usepackage{mathrsfs}
	\usepackage{amssymb}
	\usepackage{graphicx} 
	\usepackage{makeidx}
	\usepackage{amsfonts}
	\usepackage[ansinew]{inputenc}
	\usepackage[usenames, dvipsnames]{pstricks}
	\usepackage{epsfig}
	\usepackage{epstopdf}
	\usepackage{pst-grad} 
	\usepackage{pst-plot} 
	\usepackage[colorlinks, hyperindex]{hyperref}
	\colorlet{Mycolor1}{yellow!10!black!90!}
	\hypersetup
	{
		colorlinks, %
		citecolor= Red, %
		linkcolor=blue, %
		urlcolor=WildStrawberry, %
	}

\makeindex

\begin{document}

\title{\textcolor{WildStrawberry}{{\Large\bf  Quantized Brans Dicke Theory: Phase Transition and Strong Coupling (Large $\omega$) Limit \& General Relativity}}}

\author{Sridip Pal} 
\email{sridippaliiser@gmail.com; srpal@ucsd.edu}
\affiliation{Department of Physics;
University of California, San Diego\\
9500 Gilman Drive, La Jolla, California, 92093, USA}

\begin{abstract}
We show that Friedmann-Robertson-Walker (FRW) geometry with a flat spatial section in quantized (Wheeler deWitt quantization) Brans Dicke (BD)  theory reveals a rich phase structure owing to anomalous breaking of a classical symmetry, which maps the scale factor $a\mapsto\lambda a$ for some constant $\lambda$. In the weak coupling ($\omega$) limit, the theory goes from a symmetry preserving phase to a broken phase. The existence of a phase boundary is an obstruction to another classical symmetry [see \htmladdnormallink{arXiv:gr-qc/9902083}{arXiv:gr-qc/9902083}] (which relates two BD theories with different coupling) admitted by BD theory with scale invariant matter content i.e $T^{\mu}{}_{\mu}=0$. Classically, this prohibits the BD theory from reducing to General Relativity (GR) for scale invariant matter content.  We show that a strong coupling limit of both BD and GR preserves the symmetry involving the scale factor. We also show that with a scale invariant matter content (radiation i.e $P=\frac{1}{3}\rho$), the quantized BD theory does reduce to GR as $\omega\rightarrow\infty$, which is in sharp contrast to classical behavior. This is a first known illustration of a scenario, where quantized BD theory provides an example of anomalous symmetry breaking and resulting binary phase structure. We make a conjecture regarding the strong coupling limit of the BD theory in a generic scenario.
\end{abstract}

\maketitle

\section{Introduction}

Brans Dicke (BD) theory \cite{Will} is one of the closest cousins of General Relativity (GR). The salient feature of BD theory is that the curvature of geometry is nonminimally coupled with a scalar field, which makes Newton's constant $G$ a space-time dependent quantity. The significance of BD theory lies in the fact that it provides us with a simple prototype example of more realistic, sophisticated and physically motivated models including a wide class of scalar-tensor theories, having an interesting application in inflationary scenario \cite{johri,La,AM, Kolb, PJ, PJ2, AR}, and constructing potential dark energy models \cite{Amen}. Furthermore, the nonminimal coupling appears in the context of superstring theory \cite{Witten} as a low energy effective action for the dilaton-gravity sector in supergravity, as well as in  Kaluza-Klein theory \cite{TE} and DGP theory \cite{Dvali}, where the extra scalar field of the theory emerges naturally from the compactification of an extra dimension  \cite{cho}. It also appears in Galileon theories \cite{Nicolis}, proposed to explain cosmic acceleration while bypassing the Solar System constraints. To add to the list, BD theory can also be thought of as a limit of Horndeski theories \cite{h1,h2}. The further motivation and pertinence of the work that follows comes from the basic expectation that any quantum formulation of gravity requires ingredients foreign to GR, like higher order curvature correction, nonminimal coupling to matter. All of these make it meaningful to investigate scalar tensor theories as a quantum cosmological model, and because of its simplicity, BD theory is the most natural platform to explore such a quantum scenario to shed light on a wide class of scalar-tensor theories.

It is widely believed that as coupling $\omega$ becomes stronger, BD theory reduces to GR \cite{KN, AAJ, JD, JP, JD2}. In fact, this forms the basis to set lower limits of the $\omega$ parameter in Solar System experiments\cite{Will}. Albeit, there are counterexamples of several exact solutions not reducing to GR upon $\omega\rightarrow\infty$ \cite{mastuda, romero, romero1,romero2,romero3, FM1, FM2, MA} and counterarguments for nonconvergence with a scale invariant matter content, i.e, with $T^{\mu}{}_{\mu}=0$ \cite{oldnb, Faroni}. Hence, if we can show that in a quantized version, BD does reduce to GR, it would be of utmost importance. The first obstacle in this regard is that we do not have a complete picture of quantum gravity. Nonetheless, there has been recent rejuvenation in the Wheeler deWitt quantization \cite{deWitt, Wheeler} process of GR in a series of papers \cite{sp1,sp2,sp3,sp4,sp5,sachin}, where we build an effective quantum mechanical version of cosmological models. Given this resurgence in the Wheeler deWitt quantization process, it appears pertinent to explore the strong coupling limit of quantized BD using the Wheeler de Witt quantization process and to aim to answer the question posed in this formalism. In fact, there has been recent work regarding quantized BD theory \cite{nb, sachin1}.

In this article, we show for the first time that quantized BD theory can provide an elegant example of anomalous symmetry breaking leading to the existence of a rich phase structure, and thus the appeal of this work lies beyond quantum cosmology. Not to mention, the anomalous symmetry breaking is a widespread phenomenon in quantum systems ranging from particle physics to critical phenomenon in condensed matter physics, for example, relativistic quantum field theories admit chiral anomaly and weyl anomaly. In fact, the anomaly cancellation is an important tool to study quantum field theory in general.  It is known in condensed matter that a 3-body problem with a large scattering length admits Efimov states \cite{efimov} due to anomalous breaking of scale symmetry of inverse square potential down to a discrete scaling group and the resulting appearance of a limit cycle in renormalisation group (RG) flow. Generically, in a singular potential like inverse square, renormalization is required to tame the singularity near the origin. We find similar singular potential in the quantum cosmological description of BD theory where the singularity appears owing to big bang singularity. Thus, the purpose of the communication is twofold, first to provide yet another physical scenario to the list of examples ranging from superconductivity\cite{cl4,cl5}, discrete Hamiltonian models\cite{wilson1,wilson2}, quantum field theory models\cite{cl3} to {\it S}-matrix models\cite{cl1,cl2}, where limit cycle and anomalous behavior with such rich physics can be realized. On the other hand, it is expected to elucidate the quantum behavior of scalar-tensor theories in the quantum cosmological set up, specifically to show the BD theory with scale invariant matter does reduce to GR in the large $\omega$ limit. To be specific, we will study the quantized Friedmann-Robertson-Walker (FRW) metric in BD theory with radiation like matter content, having conformal invariance. It deserves mention that the conformal properties of BD theory have been studied classically \cite{valero2} as well as in the loop quantized version \cite{MY}, but such an existence of the phase structure remains to be explored. Furthermore, such novel physics has never before been reported or emphasized in the context of quantum cosmology to the best of our knowledge. 

The FRW model with a flat spatial section has a symmetry under scaling of ``scale factor" in GR. Under the scaling $a\mapsto\lambda a$, the Einstein equation of motion remains invariant. This symmetry is present in BD theory as well with a homogeneous scalar field.  In this work, we show that the symmetry does not survive the quantization process in BD theory. For some range of coupling, the symmetry is broken anomalously solely due to quantum effects, and this leads to a binary-phase structure of quantized BD theory. We will show that the strong coupling ($\omega\rightarrow\infty$) limit of BD theory is in a symmetry preserving phase and so is the quantized GR.  We argue that quantum mechanically, the presence of a phase wall must be an obstacle to the classical argument showing BD does not reduce to GR for scale invariant matter. In fact, exploiting the symmetry we explicitly show that BD theory does reduce to GR in strong coupling limit for a FRW universe with a flat spatial section and radiation (scale invariant) matter content, which is in sharp contrast with classical behavior. This contrasting behavior along with the existence of a rich quantum phenomenon should initiate more research exploring quantum BD theory along with other scalar-tensor theories, its strong coupling limit in a generic scenario.

\section{Brans Dicke Theory}

The BD theory in the Jordan frame with a perfect fluid ($P=\alpha\rho$) is described by the following Lagrangian:
\begin{equation}\label{lag}
\mathcal{L}= \phi R - \frac{\omega}{\phi}\partial_{\mu}\phi\partial^{\mu}\phi + \alpha\rho,
\end{equation}
where the scalar field $\phi$ is manifestly nonminimally coupled with the Ricci scalar. 

The line element of the FRW universe with a flat spatial slice is given by
\begin{equation}\label{metric}
ds^{2}=-n^{2}dt^{2} + a^{2}(t)\left[dx^{2}+dy^{2}+dz^{2}\right]\ .
\end{equation}
where $n^{2}(t)$ is the lapse function and $a(t)$ is the scale factor. \\

We parametrize the scale factor and $\phi$ in the following way: $a(t) = e^{\kappa(t)};\ \phi(t)= e^{\gamma(t)}\ .$ Since, we have assumed an isotropic homogeneous universe, it is only natural to assume that $\phi$ is a function of time only. Now, we  define a new variable $\beta(t)\equiv \kappa(t)+\frac{\gamma(t)}{2}$ and trade it in against $\kappa$ (as we will see this redefinition allows us to write the Lagrangian in a nice manner where $\beta$ and $\gamma$ gets decoupled, otherwise, we would have terms like $\dot{\kappa}\dot{\gamma}$).\\

Using this parametrization, the Lagrangian for the gravity sector can be written as
\begin{equation}
L_{g}=\frac{e^{3\beta-\frac{\gamma}{2}}}{n}\left[-6\dot{\beta}^{2}+\frac{2\omega+3}{2}\dot{\gamma}^{2}\right]\ .
\end{equation}

The corresponding Hamiltonian is given by
\begin{equation}\label{3.91}
H_{g}= ne^{\frac{\gamma}{2}-3\beta}\left(-\frac{p_{\beta}^{2}}{24}+\frac{p_{\gamma}^{2}}{2(2\omega+3)}\right).
\end{equation}

where $p_{\beta}$ and $p_{\gamma}$ are momenta conjugate to $\beta$ and $\gamma$ respectively.\\

For the matter sector, we take up a perfect fluid with $\alpha =\frac{1}{3}$ i.e radiation. Using standard thermodynamical considerations , the Hamiltonian for the matter sector is derived as
\begin{equation}
\label{3.93}
H_{f}= ne^{3(\frac{\gamma}{2}-\beta)\alpha}p_{T}= ne^{(\frac{\gamma}{2}-\beta)}p_{T},
\end{equation}
where $p_{T}$ is the momentum associated with fluid. A nice and crisp exposition of using the fluid sector to define a time variable $T$ and conjugate momentum $p_{T}$, is given in \cite{sp1}. The fact that the Hamiltonian of fluid sector turns out to be linear in $p_{T}$ facilitates writing down a Schrodinger-like equation.\\

 The Equations~\eqref{3.91} and \eqref{3.93} can be combined to yield the total Hamiltonian,
\begin{equation}
H = ne^{\frac{\gamma}{2}-\beta}\left(-\frac{e^{-2\beta}p_{\beta}^{2}}{24}+\frac{e^{-2\beta}p_{\gamma}^{2}}{2(2\omega+3)}+p_{T}\right).
\end{equation}

The operators are now ordered following the prescription as laid out in \cite{sp1,sp3}, and varying the Hamiltonian with respect to $n$ results in a Hamiltonian constraint, given by
\begin{equation}
\left(-\frac{1}{24}e^{-\beta}p_{\beta}e^{-\beta}p_{\beta}+\frac{e^{-2\beta}p_{\gamma}^{2}}{2(2\omega+3)}+p_{T}\right)=0.
\end{equation}
As we quantize the system, the operators are realized in ``position" space in the following way: $p_{\beta}\mapsto-\imath\partial_{\beta}$, $p_{\gamma}\mapsto-\imath\partial_{\gamma}$ and  $p_{T}\mapsto-\imath\partial_{T}$, leading to the Wheeler deWitt equation:
\begin{equation}\label{hc}
\left(\frac{1}{24}e^{-\beta}\partial_{\beta}e^{-\beta}\partial_{\beta}-\frac{e^{-2\beta}\partial_{\gamma}^{2}}{2(2\omega+3)}\right)\psi=\imath\partial_{T}\psi.
\end{equation}

A change of variable $\chi_{B}=e^{\beta}$ recasts this Hamiltonian constraint \eqref{hc} into

\begin{equation}
 \frac{1}{24}\frac{\partial^{2}\psi}{\partial\chi_{B}^{2}} -\frac{1}{2(2\omega+3)}\frac{1}{\chi_{B}^{2}}\frac{\partial^{2}\psi}{\partial\gamma^{2}}=\imath\frac{\partial\psi}{\partial T}.
 \end{equation}
 
We use the separation of variable technique $\psi(\gamma,\chi_{B},T)=\xi(\gamma)\varphi(\chi_{B})e^{\imath ET}$ to obtain: 
\begin{equation}\label{gamma}
\frac{\partial^{2}\xi}{\partial\gamma^{2}} = -k^{2} \xi;
\end{equation}
 with the solution given by $\xi = e^{\imath k\gamma}$, where $k$ appears due to separation of variables; subsequently, $\varphi$ satisfies
\begin{equation} \label{voila}
\frac{1}{24}\frac{\partial^{2}\varphi}{\partial\chi_{B}^{2}}+ \frac{k^{2}}{2(2\omega+3)}\frac{1}{\chi_{B}^{2}}\varphi =- E\varphi.
\end{equation}

We define parameters
\begin{equation}\label{coupling}
g= \frac{12k^{2}}{2\omega+3},\ E^{\prime}=24E,
\end{equation}
to cast Eq.~\eqref{voila} in the following form:
\begin{equation}\label{ge}
-\frac{\partial^{2}\varphi}{\partial\chi_{B}^{2}}-\frac{g}{\chi_{B}^{2}}\varphi = E^{\prime}\varphi.
\end{equation}

So, we have transformed this problem to a well-known inverse square potential problem with an attractive potential for $g>0$ i.e $\omega>-\frac{3}{2}$, repulsive one for $g<0$, i.e, $\omega<-\frac{3}{2}$. Apparently Eq.~\eqref{ge} admits a scaling symmetry under $\chi_{B}\mapsto \lambda\chi_{B}$, which is reminiscent of classical scale symmetry. To be specific, if $\phi(\chi_{B})$ is an eigenstate with energy $E^{\prime}$, then $\phi(\lambda\chi_{B})$ is an eigenstate with energy with $\lambda^{2}E^{\prime}$. This also implies a continuous spectra i.e if $E^{\prime}$ is an eigenenergy, then there exists a state with energy $\lambda^{2} E^{\prime}$ for $\lambda\in\mathbb{R}$. For $g<\frac{1}{4}$, one can show that $E^{\prime}>0$, and we have a spectra bounded below. For a strongly coupled regime, $g>\frac{1}{4}$, there exist states with negative $E^{\prime}$ which indicates that if we have to preserve scaling symmetry, there can not be any ground state. This comes out of \textcolor{WildStrawberry}{\textit{S-theorem}} elucidated nicely in the appendix of \cite{srben}. Hence, in a strongly coupled regime, we need to do a self-adjoint extension of the Hamiltonian \cite{gupta} or equivalently we need to regularize and renormalize \cite{swingle} the coupling so as to ensure a ground state. This is precisely what leads to anomalous (quantum) breaking of scale symmetry for $g>\frac{1}{4}$ \cite{camblong}. In summary, owing to quantum effects, we have two distinct phases: in the weakly attracting and repulsive regime ($g<\frac{1}{4}$) the symmetry is preserved, while in the strongly attractive regime($g>\frac{1}{4}$) the symmetry breaks down. It has been shown \cite{swingle,beane} that the symmetry is not lost completely but rather broken down to a discrete scaling symmetry, and we have limit cycle behavior in theory space. The critical point $g=\frac{1}{4}$ translates to a parabola in $(k,\omega)$ space (see Fig.~1), given by
\begin{equation}
\omega = \frac{48k^{2}-3}{2}.
\end{equation}
\begin{figure}[!ht]
 \includegraphics[scale=0.55]{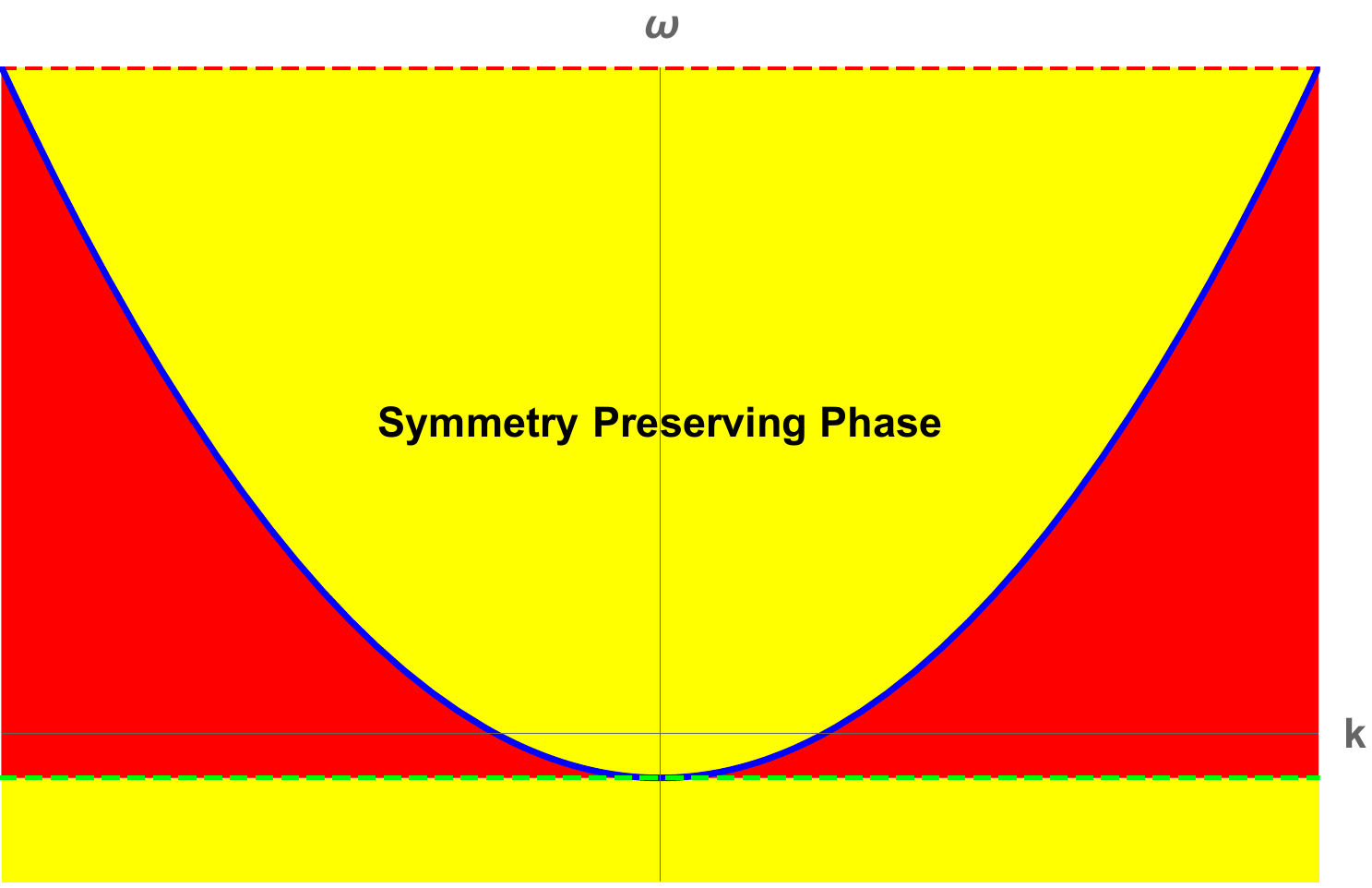}
 \caption{Phase structure in $(k,\omega)$ plane; The red (dark shaded) region is where symmetry is broken due to quantum effects while in the yellow (lightly shaded) region, the symmetry is preserved. The thick blue line represents the phase wall. The dotted red line is supposed to be at $\omega=\infty$. The dotted green line below which we have the yellow (lightly shaded) region is at $\omega=-1.5$.}
\end{figure}

where $k$ is the eigenvalue of the $p_{\gamma}$ operator, i.e, $k$ can be thought of as momentum associated with $\gamma$ and $\omega$ is coupling of the BD theory. This $k$ dependence of the critical point can be interpreted in the following way, which is very popular in  field theory community: the scalar field (hence, the system as a whole) is composed with different momentum $k$ modes, which do not talk with each other and evolve independently; just like a free field theory. Each of these modes exhibits phase transition at a critical point, which is a function of its momenta.\\

For a given coupling $\omega$ such that $2\omega +3>0$, if we are to preserve the symmetry in the quantized version, then we restrict the possible momentum modes in a range i.e $|k|<\frac{1}{4}\sqrt{\frac{2\omega+3}{3}}$.  Only in the limit $\omega\rightarrow\infty$, all the momentum modes are allowed. It is worth noting that for a fixed $\omega$, $g$ is invariant under $k\mapsto -k$. Hence, in the regime where $2\omega+3>0$, i.e, $g$ is positive definite, for $k > 0$ as well as for $k < 0$, the universe can be in either phase. But, for $2\omega+3<0$, $g$ is negative definite, i.e, $g< 0 <\frac{1}{4}$, therefore the symmetry is always preserved. The yellow region (lightly shaded) below the $\omega=\frac{-3}{2}$ horizontal line represents this regime in the graph. It also deserves mention that for a given nonzero mode $k$ such that $|k|< \frac{1}{4}$, the broken phase is attained only when $\omega$ becomes negative, to be precise when $\frac{-3}{2}<\omega <0$. Furthermore, the $k=0$ mode is very special in the sense that it never undergoes phase transition for any value of coupling $\omega$. \\

\section{Breakdown of Faraoni classical symmetry}

The BD theory with scale invariant matter content has  a classical symmetry as pointed out in \cite{Faroni,Faroni2}. Two Brans Dicke space-time $\left(M,g_{\mu\nu}^{(\omega)},\phi^{(\omega)}\right)$ and $\left(M,\tilde{g}_{\mu\nu}^{\tilde{\omega}},\tilde{\phi}^{\tilde{\omega}}\right)$ are equivalent if we have $\tilde{\phi}=\phi^{1-2\theta} \Leftrightarrow \tilde{\gamma}=\gamma\left(1-2\theta\right)$, $\tilde{g}_{\mu\nu}= \phi^{2\theta}g_{\mu\nu} \Leftrightarrow \tilde{\beta}=\beta$ and 
$\tilde{\omega}= \frac{\omega +6\theta(1-\theta)}{(2\theta-1)^{2}}$.\\

This symmetry is Abelian in nature and described by one parameter $\theta$. By this mapping i.e choosing $\theta$ suitably, we can classically relate two $\omega$ across a phase transition. In fact, $\omega\rightarrow\infty$ can be thought of as moving  within this equivalence class. Now GR does not have this classical symmetry, implying GR cannot belong to this equivalence class. Thus GR cannot be classically realized as a strong coupling limit of BD theory with scale invariant matter content. Nonetheless, in the quantized version, the $\omega\rightarrow\infty$ limit of BD theory always lies in a symmetry preserving phase. Had this symmetry been there quantum mechanically, we could choose $\theta$ aptly [$\theta = \frac{1}{2} \left(1\pm \sqrt{\frac{\omega_{ns}+\frac{3}{2}}{\omega_{s}+\frac{3}{2}}}\right)$] to approach the limit and conclude that a theory in a broken phase with $\omega_{ns}$  is equivalent to a theory in a symmetry preserving phase with $\omega_{s}>\omega_{ns}\geq\frac{-3}{2}$. But quantum mechanically the nature of the spectrum changes dramatically across the phase transition. Thus this classical sense of equivalence must break down quantum mechanically and so must the argument proving that the GR is not a strong coupling limit of BD with $T^{\mu}{}_{\mu}=0$. \\

One can modify the argument by Faraoni and argue that within the symmetric ($a\mapsto\lambda a$) phase, there is no phase wall, hence, the classical Faraoni equivalence might survive in this phase. The $\omega\rightarrow\infty$ limit is in this symmetry preserving phase, and hence lies in the Faraoni equivalence class. This modified (restricted) sense of equivalence has no obstruction coming from the phase transition wall. Albeit, as we will show below, the strong coupling limit of BD does reduce to GR for a FRW metric with a flat spatial slice and radiation like matter content. 

\section{Strong coupling limit and GR}
In this section, we will explicitly probe the strong coupling limit of BD and compare it to GR in the quantized version. The FRW line element is again given by Eq.~\eqref{metric} and we parametrize $a=e^{\sigma(t)}$.\\

The fluid sector can be dealt with in a similar manner as in BD, following the operator ordering prescription to arrive at the Hamiltonian of quantized GR
\begin{equation}
\hat{H}=ne^{3\alpha\sigma}\left(\frac{1}{24}e^{-\frac{3(1-\alpha)}{2}\sigma}\partial_{\sigma}e^{-3\frac{(1-\alpha)}{2}\sigma}\partial_{\sigma}+p_{T}\right),
\end{equation}
and a change of variable for $\alpha=\frac{1}{3}\neq 1$, $\chi_{G}=Exp\left[\frac{3(1-\alpha)}{2}\sigma\right]=Exp[\sigma]$
recasts the Wheeler de Witt equation $\hat{H}\Psi =0$ into $\frac{1}{24}\frac{\partial^{2}\Psi}{\partial\chi_{G}^{2}}=\imath\partial_{T}\Psi$. Plugging in the ansatz $\Psi=\psi(\chi_{G})e^{\imath E T}$, we obtain 
\begin{equation}\label{ge2}
-\frac{1}{24}\frac{\partial^{2}\psi}{\partial\chi_{G}^{2}}= E\psi .
\end{equation}

This precisely mimics the $g\rightarrow 0$ limit of BD theory as in this limit the governing equation \eqref{ge} becomes
\begin{equation}\label{ge1}
-\frac{1}{24}\frac{\partial^{2}\varphi}{\partial\chi_{B}^{2}}= \frac{1}{24}E^{\prime}\varphi=E\varphi.
\end{equation}

Thus governing equations~ \eqref{ge1} and~\eqref{ge2}, controlling the behavior of $\chi_{B}$ and $\chi_{G}$ are same. In fact both of them admit symmetry under scaling of $\chi_B$ and $\chi_{G}$; albeit the scale factor behaves differently in these two scenarios. In GR, the scale factor $a$ is given by $a=\chi_{G}$ while in BD theory, it is given by $a= e^{-\frac{\gamma}{2}}\chi_{B}$.\\

Now, for $g\neq0$, $\varphi(\chi_{B})$ depends on $g$ (the solution being given by the modified Bessel function of order $\sqrt{-g+\frac{1}{4}}$), and hence on momentum mode $k$ \eqref{coupling} of scalar field $\gamma$ \eqref{gamma}. As $\omega\rightarrow\infty$, $g$ becomes $0$ and this dependence goes away. Even if we make a time dependent state by superposing energy eigenfunctions $\varphi$, the behavior of $\gamma$ is unaffected. On the other hand, even if we superimpose various momentum modes of $\gamma$, that does not affect the evolution of $\varphi$. Hence, in the $\omega\rightarrow\infty$ limit, the wave function $\xi(\gamma)$ controlling the behavior of $\gamma$ is explicitly time independent, which, in turn implies that on expectation value level, the GR FRW thus obtained has a scale factor that is some time independent multiple of the scale factor obtained from the strong coupling limit of BD. Thus for some constant $c$, we can write $\langle a_{GR} \rangle=c \langle a_{BD}\rangle$.\\

We know the strong coupling limit of both BD and GR preserves symmetry even after quantization; hence $\langle a_{GR}\rangle $ and $\langle a_{BD}\rangle $ are related by symmetry transformation. Thus, we have been able to show that quantum FRW obtained from BD does reduce to quantum FRW obtained from GR.  For example, by superposing solutions of \eqref{gamma}, one can have $\xi(\gamma)=\frac{1}{\sqrt[4]{2\pi^{3}}} \int dk\ e^{-k^{2}+ik\gamma}=\frac{1}{\sqrt[4]{2\pi}}e^{-\frac{\gamma^{2}}{4}}$, to obtain $c=\langle e^{-\frac{\gamma}{2}}\rangle =e^{\frac{1}{8}}$. One might wonder about the fluctuation of $\gamma$, but note, in the strong coupling limit, even the fluctuations are time independent. Hence, even in the sense of the operator, we have $a_{GR}=\mathcal{C}. a_{BD}$ for constant operator $\mathcal{C}$. For example, $\sqrt{\langle \mathcal{C}^{2}\rangle -\langle \mathcal{C}\rangle^{2} } = e^{\frac{1}{8}}\sqrt{e^{\frac{1}{4}}-1}$ for the above mentioned $\xi$.\\

\section{D\'{e}nouement}

We have shown the existence of a binary phase structure of the FRW model with a flat spatial section in quantized BD theory, identifying the phase transition wall, explaining how the quantum effects break the classical symmetry which maps $a\mapsto\lambda a$. The obstruction provided by the phase transition wall implies the argument, showing that the BD theory with a scale invariant matter content does not reduce to GR and does not go through in the quantized version. Hence, we explore the strong coupling limit of the quantized BD theory and show explicitly that in sharp contrast with classical behavior, quantum mechanically, it does reduce to GR for a scale invariant matter content i.e radiation. This result is of utmost importance considering the fact that Solar System experiments and various important aspects of BD theory underlie the assumption that in the large $\omega$ limit, BD reduces to GR.\\

Although we have been working with the FRW model, it is a straightforward but nonetheless exciting exercise to show that the anisotropic homogeneous Bianchi-I model exhibits such scaling symmetry at the classical level which breaks down at the quantum level for a region in coupling space. Unlike FRW, Bianchi-I exhibits such binary phase structure in both GR and BD theories. We wish to report on it in future. \\

The invariance under $a\mapsto\lambda a$ plays a role in showing the convergence of strongly coupled BD to GR in quantized version. Hence, it seems that in the generic scenario, the strong coupling limit of the quantized BD theory yields a space-time, whose spatial slice (upon ADM decomposition) is conformal to the spatial slice of space-time obtained from quantized GR. At present, this is merely a conjecture, requiring a rigorous proof to be established. Nonetheless, this seems quite natural, as in the Einstein frame description of the BD theory, the scalar field always gets decoupled. There will possibly be a way to establish this decoupling effect in the Jordan frame or, to be more ambitious, to prove an equivalence between Jordan and Einstein frame descriptions of the BD theory in a generic scenario.\\

Last but not least, we list open questions that we believe will be interesting to explore in future: 
\begin{enumerate}
\item  to investigate whether the symmetry as laid out by Faraoni breaks down quantum mechanically, in a generic scenario or it happens only in FRW with a flat spatial section. One obvious choice would be to explore FRW with a curved spatial slice. 
\item to explore the strong coupling limit of the BD theory and issue of convergence to GR in a generic scenario in the quantized version. One can investigate a generic scalar-tensor theory in a similar setup.
\item to explore whether any other model in quantized BD exhibits such rich quantum physics like anomalous symmetry breaking.
\item to show (in)equivalence of Einstein and Jordan frames with matter content.
\item to investigate the cosmological implication of anomalous symmetry breaking in the FRW model.
\item for the loop quantum gravity community to test whether the result obtained is robust enough to be independent of the quantization scheme and to be found in the loop quantum cosmological setup as well even though the work above has been done in a mini-superspace quantization scheme.
\end{enumerate}

\textit{Note Added}: A week after this had been posted in the arXiv, a work \cite{fabris} regarding self-adjoint extension in Brans-Dicke has appeared, where they arrived at a similar singular potential and found a constraint on operator ordering to ensure self-adjointness. It deserves mention that in the context of a singular potential, self-adjoint extension and renormalisation is intricately related. Hence, the results of \cite{fabris} can potentially be translated in the language of renormalisation and anomalous breaking of scale symmetry. They obtained an inequality involving momentum of scalar field and a parameter that depends on the operator ordering, coupling $\omega$, which ensures that the Hamiltonian is essentially self-adjoint.  The regime of coupling where the Hamiltonian is essentially self-adjoint is precisely the regime where the symmetry is preserved whereas in the complementary regime, the symmetry breaks anomalously.\\

\begin{acknowledgments}
The author would like to thank Narayan Banerjee for insightful comments on the manuscript. The author also thanks anonymous referee for pointing out subtlety in the graphical representation of phase structure. This work was supported in part by the U.S. Department of Energy under Contract No. DE- SC0009919.
\end{acknowledgments}

{}
\end{document}